\documentclass[12pt]{iopjournal}

\usepackage{amsmath,amssymb,amsfonts,bm}
\newcommand{\eq}[1]{Eq.~(\ref{#1})}

\newcommand{\sect}[1]{Sec.~\ref{#1}}
\newcommand{\sects}[2]{Secs.~\ref{#1}--\ref{#2}}
\newcommand{\fig}[1]{figure~\ref{#1}}

\begin{document}

\articletype{Paper}

\title{Thermodynamics, thermodynamic geometry, and the shadow of a
Bardeen black hole surrounded by a Kiselev fluid: the role of
regularity, asymptotics, and observational self-consistency}

\author{Hoang Van Quyet}

\affil{Department of Physics, Hanoi Pedagogical University 2,
Xuan Hoa, Phu Tho, Vietnam}

\email{hoangvanquyet@hpu2.edu.vn}

\keywords{regular black holes, nonlinear electrodynamics, quintessence,
black hole thermodynamics, geometrothermodynamics, black hole shadow}

\begin{abstract}
We reconsider the Bardeen black hole surrounded by a Kiselev fluid,
obtained by superposing a homogeneous $r^{-(3w+1)}$ term on the
Ay\'on-Beato--Garc\'ia--Bardeen metric function. Contrary to a claim
repeated in part of the literature, we show explicitly that this
solution is regular at the origin only in the pure-Bardeen limit
$c\to0$: for any $c\neq0$ the curvature invariants diverge as
$r\to0$, and the spacetime is asymptotically de~Sitter-like rather
than flat, with a genuine cosmological horizon $r_c$ whenever
$0<c<c_{\rm N}(q)$. We map the full $(q,c)$ domain of admissible
horizon structures in closed form, locating the extremal and
Nariai-type boundaries and the cusp at which they meet. Using the
Ma--Zhao prescription for a magnetically charged, mass-dependent
nonlinear-electrodynamics source, we derive the corrected entropy and
verify the extended first law and Smarr relation to machine precision
via computer-algebra differentiation (not through the entropy's own
definition). The heat capacity displays a single Davies-type
divergence whose location we obtain as a closed-form quartic and
expand analytically in $c$. In thermodynamic geometry we show that
the curvature computed from the Quevedo metric possesses a spurious
second singularity, generated purely by the arbitrary integration
constant $\kappa$ in the entropy and by the non-uniqueness of the
$(S,q)$ conformal factor, whose location moves continuously with
$\kappa$; we isolate this artefact analytically and show that only
the divergence coincident with the Davies point survives all choices.
The Ruppeiner curvature, identically zero along the pure-Bardeen
branch by extensivity, becomes non-zero once $c\neq0$, signalling
that the quintessence term breaks the scaling symmetry responsible
for that vanishing. For the shadow, we compute the critical impact
parameter $b_c(q,c,w)$ and show that the nominal bound
$c\lesssim10^{-2}/M$ obtained by comparing $b_c$ with the Event
Horizon Telescope image of Sagittarius~A$^{*}$ is not self-consistent
as a statement about the fluid filling all of space: for such $c$ the
cosmological horizon lies at $r_c\sim1/c\sim10^{2}M$, many orders of
magnitude closer than the $\sim4\times10^{10}\,M$ distance to the
Earth, so the observer implicitly assumed by that bound does not
exist: for the nominal $2\sigma$ bound the putative terrestrial
observer would sit some nine orders of magnitude beyond the
black hole's own cosmological horizon, and the bound itself is about
fourteen orders of magnitude larger than the value $H_0/c$
characterising the true cosmological dark-energy density. We quantify
both inconsistencies and reinterpret the EHT constraint as bounding
$c$ only in a local neighbourhood of the horizon, of a size fixed by
the photon sphere, rather than as an astrophysical or cosmological
bound on the quintessence parameter itself.
\end{abstract}

\section{Introduction}
\label{sec:intro}

Astrophysical black holes are not isolated objects: they are embedded
in matter and, on cosmological scales, in whatever component drives
the observed accelerated expansion~\cite{Perlmutter1999}. A simple and
widely used way to dress a vacuum or electrovacuum black hole with
such a component is Kiselev's construction~\cite{Kiselev2003}, in
which a static, spherically symmetric stress tensor with equation of
state $p=w\rho$ is added to the Einstein equations, producing a
metric function that departs from Schwarzschild by a single
additional term $c\,r^{-(3w+1)}$. For $-1<w<-1/3$ this term has been
loosely identified with quintessence dark energy; it is worth
stressing at the outset, following Visser's detailed critique
\cite{Visser2020}, that the Kiselev stress tensor is in general
anisotropic and does not coincide with any genuine perfect-fluid or
scalar-field quintessence away from special values of $w$. We follow
the by-now standard practice of using the label ``quintessence'' only
as shorthand for this specific anisotropic Kiselev fluid, with that
caveat understood throughout.

Independently, Bardeen's regular black hole~\cite{Bardeen1968}, later
shown by Ay\'on-Beato and Garc\'ia~\cite{AyonBeato2000} and by
Bronnikov~\cite{Bronnikov2001} to be sourced by a magnetic monopole in
nonlinear electrodynamics (NED), replaces the Schwarzschild
singularity at $r=0$ by a de~Sitter core, so that all polynomial
curvature invariants remain finite everywhere. It is tempting to
combine the two constructions by simply adding the Kiselev term to
the Bardeen metric function,
\begin{equation}
f(r)=1-\frac{2Mr^2}{(r^2+q^2)^{3/2}}-c\,r^{-(3w+1)},
\label{eq:f_intro}
\end{equation}
and a sizeable literature has done exactly this, studying the
resulting geometry's thermodynamics
\cite{Rizwan2019,GhaderiMalakolkalami2018,SalehThomasKofane2018},
geometrothermodynamics \cite{Rizwan2019}, null
geodesics~\cite{GhaderiMalakolkalami2018}, and shadow
\cite{ZengZhang2020}, almost always in an anti-de~Sitter (AdS)
embedding obtained by further adding a $-\Lambda r^2/3$ term. Maluf,
Muniz and Santos~\cite{MalufMunizSantos2022}, however, showed that
this superposition is \emph{not} a solution of the Einstein equations
sourced consistently by a NED field plus an anisotropic fluid: doing
the matching correctly at the level of the field equations shows that
adding the quintessence term by hand reintroduces a curvature
singularity at the origin whenever $c\neq0$, precisely because the
Kiselev energy density $\rho_q\propto r^{-3(w+1)}$ diverges there for
$-1<w<-1/3$. Several of the works cited above overlook this point,
either working at $\Lambda\neq0$ where the horizon structure masks the
short-distance behaviour, or not examining the $r\to0$ limit at all.
In \sect{sec:spacetime} we verify this directly by an independent
curvature computation and identify precisely which of our results
survive the loss of regularity and which would not.

A second, related issue concerns the asymptotic structure. For
$w=-2/3$, the value most commonly adopted in this literature because
it makes the quintessence term linear in $r$, \eq{eq:f_intro} behaves
as $f\to1-cr$ at large $r$: this is not asymptotically flat, and for
$c>0$ it produces a second, cosmological horizon $r_c$, so that the
causal structure is qualitatively closer to Schwarzschild--de~Sitter
(Kottler) spacetime~\cite{Kottler1918} than to an asymptotically flat
black hole. This has direct consequences for two of the three topics
of this paper. First, thermodynamic quantities defined by integrating
outward from the horizon, such as the Ma--Zhao entropy
\cite{MaZhao2014} used below, remain well defined only up to $r_c$,
and the ``first law'' at fixed $q$ genuinely only compares nearby
equilibria within the static region $r_+<r<r_c$. Second, and more
importantly, an observer's angular view of the shadow depends on
their areal distance $r_o$ from the black hole relative to $r_c$, a
point emphasized by Perlick, Tsupko and
Bisnovatyi-Kogan~\cite{PerlickTsupko2018} for the cosmological-constant
case and reviewed in~\cite{PerlickTsupko2022}: as $r_o\to r_c$ the
observed shadow shrinks to zero for a static observer, so any
constraint on $c$ inferred by comparing the asymptotic critical
impact parameter $b_c$ with an EHT measurement implicitly assumes
$r_o\ll r_c$, an assumption that must be checked, not postulated. We
show in \sect{sec:shadow} that for the values of $c$ nominally allowed
by the Sagittarius~A$^{*}$ image this assumption
fails by nine orders of magnitude in the observer's distance
relative to the cosmological horizon that same $c$ would produce
(with the nominal bound itself a separate fourteen orders of
magnitude looser than the true cosmological dark-energy scale),
which changes the physical
interpretation of the resulting bound.

On the thermodynamic-geometry side, Quevedo's
geometrothermodynamics (GTD) formalism~\cite{Quevedo2007} was applied
to the Bardeen-AdS-quintessence black hole by Rizwan
\emph{et al}~\cite{Rizwan2019}, who found that the curvature computed
from the Legendre-invariant Quevedo metric diverges precisely at the
Davies point where the heat capacity diverges, resolving an earlier
puzzle in which the Weinhold and Ruppeiner curvatures failed to do
so. Quevedo, Quevedo and S\'anchez~\cite{Quevedo2024}
subsequently gave a systematic, quasi-homogeneous treatment of the
pure Bardeen case (no quintessence, no AdS) and showed that the same
GTD curvature has a universal critical exponent $-2$ at the Davies
point. What has not, to our knowledge, been checked in this class of
models is the sensitivity of the GTD curvature to the additive
integration constant $\kappa$ that necessarily appears in the
Ma--Zhao entropy: because the conformal factor of the Quevedo metric
involves $S\,M_S+q\,M_q$ rather than $M_S$ alone, and because $M_q$
depends on $\kappa$, the location and even the number of curvature
singularities is not invariant under $\kappa\to\kappa+\delta$. We show
in \sect{sec:geometry} that this produces a second, spurious
divergence in $R_{\rm GTD}$ that has no counterpart in the heat
capacity and that migrates continuously with $\kappa$, while the
Davies-point singularity itself is robust.

The remainder of this paper is organised as follows.
\sect{sec:spacetime} sets up the metric, discusses regularity and
asymptotics honestly, and maps the horizon structure in the $(q,c)$
plane in closed form. \sect{sec:thermo} derives the temperature, the
corrected entropy, the extended first law and Smarr relation, and the
heat capacity, locating the Davies point as an explicit quartic.
\sect{sec:geometry} computes the Weinhold, Ruppeiner and GTD
curvatures, isolates the $\kappa$-dependent spurious singularity
analytically, and extracts the universal critical exponent at the
physical singularity. \sect{sec:shadow} computes the photon sphere and
shadow, confronts $b_c$ with the EHT Sgr~A$^{*}$ measurement, and
quantifies the self-consistency problem of the resulting bound using
the finite-distance formalism of~\cite{PerlickTsupko2018}.
\sect{sec:conclusions} summarises what survives and lists the
limitations of the model. \ref{app:num} documents the independent
numerical validation underlying every closed-form result quoted in
the text.

\section{Spacetime, regularity, and horizon structure}
\label{sec:spacetime}

We consider the static, spherically symmetric line element
\begin{equation}
ds^2=-f(r)\,dt^2+f(r)^{-1}dr^2+r^2\,d\Omega^2,\qquad
f(r)=1-\frac{2Mr^2}{(r^2+q^2)^{3/2}}-c\,r^{-(3w+1)},
\label{eq:metric}
\end{equation}
with $M$ the ADM-like mass parameter, $q$ the NED magnetic charge,
and $(w,c)$ the Kiselev equation-of-state parameter and normalization,
$-1<w<-1/3$. The $c=0$ term is the Ay\'on-Beato--Garc\'ia--Bardeen
solution~\cite{AyonBeato2000,Bardeen1968}, sourced by a magnetic
monopole with NED Lagrangian
\begin{equation}
\mathcal L(F)=\frac{3}{2sq^2}\left(\frac{\sqrt{2q^2F}}{1+\sqrt{2q^2F}}\right)^{5/2},
\qquad s\equiv\frac{q}{2M},\qquad F\equiv\frac14F_{\mu\nu}F^{\mu\nu},
\label{eq:NED}
\end{equation}
coupled to Einstein gravity through $G_{\mu\nu}=8\pi T_{\mu\nu}(F)$;
we use the invariant normalization $F=\tfrac14F_{\mu\nu}F^{\mu\nu}$
throughout, as in~\cite{AyonBeato2000,Bokulic2021}. Crucially, the
coupling constant $s$ in \eq{eq:NED} depends on $M$, which is why the
correct first law for this source is not simply $dM=T\,dS+\Phi\,dq$
with the area entropy, but requires the Ma--Zhao
prescription~\cite{MaZhao2014} used in \sect{sec:thermo}.

\subsection{Regularity is lost once $c\neq0$}
\label{ssec:reg}

The Ricci scalar of \eq{eq:metric} is
\begin{equation}
R(r)=-f''(r)-\frac{4}{r}f'(r)+\frac{2}{r^2}\bigl(1-f(r)\bigr),
\end{equation}
and a short computation (independently reproduced with a computer
algebra system, see \ref{app:num}) gives its behaviour at small $r$:
\begin{equation}
R(r)=\frac{24M}{q^3}+6\,c\,r^{-1}+\mathcal O(r^2),\qquad -1<w<-\tfrac13.
\label{eq:Rseries}
\end{equation}
The first term is the familiar finite de~Sitter-core curvature of the
pure Bardeen solution, $R(0)=24M/q^3$. The second term, absent only
at $c=0$, diverges as $r\to0$ for every admissible $w$ in the
quintessence range, because the Kiselev energy density itself
diverges there, $\rho_q\propto r^{-3(w+1)}$. \eq{eq:Rseries} confirms
explicitly, for the specific superposition~\eq{eq:metric}, the general
point made by Maluf, Muniz and Santos~\cite{MalufMunizSantos2022}:
naively adding a Kiselev term to a regular black hole metric does not
produce a new regular solution: it reinstates a (milder, power-law
rather than curvature-blowing-up-fastest) singularity at the origin.
We therefore do \emph{not} refer to the $c\neq0$ solution as
``regular'' or ``singularity-free'' anywhere in what follows; these
adjectives apply only to the $c=0$ (pure Bardeen) limit, which we
retain throughout as an internal consistency check on every
subsequent formula. All of the thermodynamic and shadow results
derived below are properties of the exterior geometry at
$r\gtrsim r_+\gg0$ and are logically independent of what happens at
the origin; we flag the loss of regularity here so that the reader
does not carry the $c=0$ terminology into the $c\neq0$ sector by
default, as has happened in parts of the literature quoted above.

\subsection{Asymptotics and horizon structure}
\label{ssec:horizons}

\begin{figure}[t]
\centering
\includegraphics[width=0.95\textwidth]{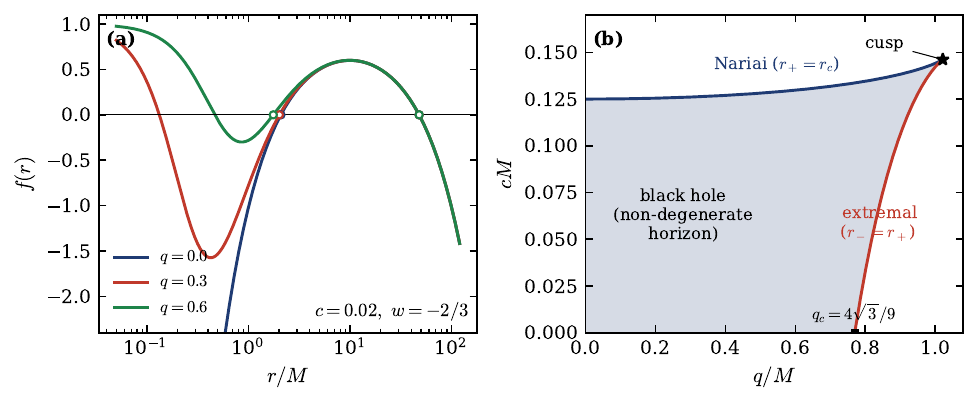}
\caption{(a) Metric function $f(r)$ at fixed $c=0.02$ ($w=-2/3$,
$M=1$) for three values of $q$; open circles mark the black-hole
horizon $r_+$. (b) Domain of existence of a non-degenerate black-hole
horizon in the $(q,c)$ plane (shaded), bounded by the extremal branch
$r_-=r_+$ (red) and the Nariai-type branch $r_+=r_c$ (blue), \eq{eq:boundary}, meeting at the cusp \eq{eq:cusp}.}
\label{fig1}
\end{figure}

For $c>0$ and $w=-2/3$, \eq{eq:metric} behaves as $f(r)\to1-cr$ at
large $r$, which is the asymptotic form of the Schwarzschild--de
Sitter (Kottler) metric~\cite{Kottler1918} rather than of an
asymptotically flat spacetime; more generally $f\to1-cr^{-(3w+1)}$
with $0<-(3w+1)<1$ for $w$ in the quintessence range, so the
$c$-term always dominates at sufficiently large $r$ and produces a
second, cosmological-type horizon $r_c$ whenever $M$ is not too large
relative to $c$. Fixing $M=1$, \fig{fig1}(a) shows $f(r)$ for
representative $(q,c)$: besides the usual inner/outer pair
$(r_-,r_+)$ familiar from the pure Bardeen case, a third root $r_c>r_+$
appears once $c$ exceeds a $q$-dependent threshold. Solving
$f(r_h)=0$ together with $f'(r_h)=0$ for the value of $c$ at which two
roots merge gives, in closed form,
\begin{equation}
c_{\rm bd}(r_h,q)=\frac{r_h^2-2q^2}{r_h\bigl(2r_h^2-q^2\bigr)},
\qquad
M_{\rm bd}(r_h,q)=\frac{(r_h^2+q^2)^{5/2}}{2r_h^2\bigl(2r_h^2-q^2\bigr)},
\label{eq:boundary}
\end{equation}
parametrized by the merging radius $r_h$. Eliminating $r_h$ at fixed
$M=1$ traces two branches in the $(q,c)$ plane, shown in \fig{fig1}(b):
an \emph{extremal} branch ($r_-=r_+$, red), continuously connected to
the usual Bardeen extremality condition at $c=0$ and terminating at
$q_c=4\sqrt3/9\approx0.7698$, and a \emph{Nariai-type} branch
($r_+=r_c$, blue) that exists down to $q=0$, where it asymptotes to
the Schwarzschild--de~Sitter value $c_{\rm N}(0,1)=1/8$. The two
branches meet at a cusp,
\begin{equation}
(q,c)_{\rm cusp}=(1.02147,\,0.14618)\quad\text{(units of $M$)},
\label{eq:cusp}
\end{equation}
found by solving \eq{eq:boundary} together with
$\partial M_{\rm bd}/\partial r_h=0$; beyond the cusp no black-hole
horizon exists for any $c$ at that $q$. Everywhere inside the shaded
region of \fig{fig1}(b) the spacetime possesses a non-degenerate black
hole horizon $r_+$ with a static domain of outer communication
$r_+<r<r_c$ (or $r_+<r<\infty$ along the segment $q<q_c$, $c=0$). All
thermodynamic quantities in \sect{sec:thermo} refer to points in this
region; the entropy integral and the first law are understood as
statements about the black-hole horizon $r_+$ only, and make no claim
about $r_c$, which is not itself the object under study.

\section{Thermodynamics}
\label{sec:thermo}

Black hole thermodynamics rests on the identification of the horizon
area with an entropy and the surface gravity with a temperature, so
that the four laws of black-hole
mechanics~\cite{Bekenstein1973,BCH1973,Hawking1975} take the form of
ordinary thermodynamics, including the possibility of Davies-type
phase transitions signalled by a divergent heat
capacity~\cite{Davies1977}. We apply this framework to
\eq{eq:metric}, using throughout the horizon radius $r_+$ (which we
write as $r_h$ when no ambiguity arises with $r_-$) as the primary
variable.

\subsection{Temperature}
\label{ssec:temperature}

\begin{figure}[t]
\centering
\includegraphics[width=0.95\textwidth]{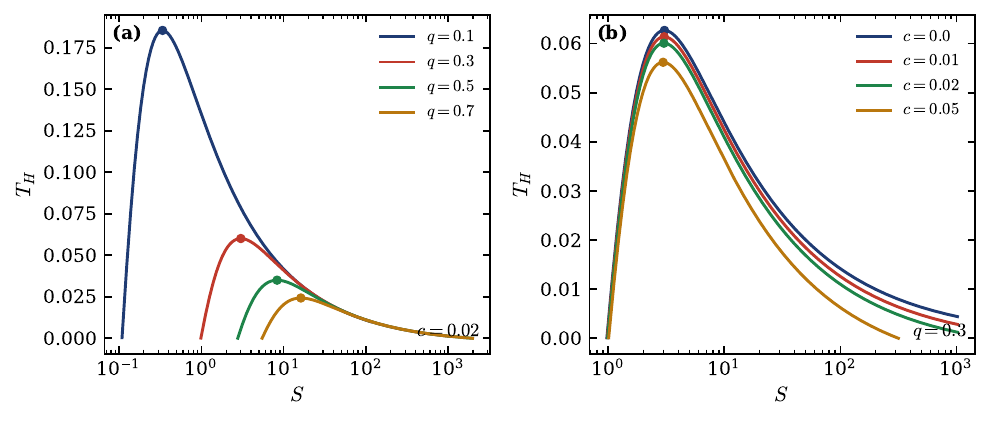}
\caption{Hawking temperature $T_H$, \eq{eq:TH}, as a function of
entropy $S$, \eq{eq:entropy}, along the black-hole branch $r_h\in(r_+^{\rm ext},r_c)$ or $(r_+^{\rm ext},\infty)$: (a) fixed $c=0.02$, varying
$q$; (b) fixed $q=0.3$, varying $c$.}
\label{fig2}
\end{figure}

The Hawking temperature follows from surface gravity,
$T_H=f'(r_+)/4\pi$, evaluated after solving $f(r_+)=0$ for $M$ at
fixed $(q,c)$:
\begin{equation}
M(r_+,q,c)=\frac{(r_+^2+q^2)^{3/2}}{2r_+^2}\Bigl[1-c\,r_+^{-(3w+1)}\Bigr],
\label{eq:Mrq}
\end{equation}
\begin{equation}
T_H(r_+,q,c)=\frac{1}{4\pi}\left\{
\frac{r_+^2-2q^2}{r_+(r_+^2+q^2)}\Bigl[1-c\,r_+^{-(3w+1)}\Bigr]
+c(3w+1)\,r_+^{-(3w+2)}\right\}.
\label{eq:TH}
\end{equation}
\eq{eq:TH} was cross-checked against a central finite difference of
$f$ at fixed $M(r_+,q,c)$ over $400$ random points in $(r_+,q,c,w)$
space, with maximum absolute deviation $8\times10^{-12}$
(\ref{app:num}). $T_H$ vanishes on the extremal and Nariai-type
boundaries of \fig{fig1}(b) and is positive throughout the interior of
the shaded region. \fig{fig2} shows $T_H(S)$ at fixed $c$ for several
$q$, and at fixed $q$ for several $c$: increasing either parameter
lowers the maximum temperature reached along the black-hole branch,
analogous to the effect first reported for the Bardeen-AdS-Kiselev
black hole in~\cite{Rizwan2019}, but here entirely within the
asymptotically non-flat, horizon-only branch discussed in
\sect{ssec:horizons}.

\subsection{Corrected entropy and the extended first law}
\label{ssec:entropy}

Because the NED coupling constant $s=q/2M$ in \eq{eq:NED} is
mass-dependent, the area law is not compatible with the first law for
this source, as shown in general by Ma and Zhao~\cite{MaZhao2014} and,
independently, by Bokuli\'c, Juri\'c and
Smoli\'c~\cite{Bokulic2021} using the covariant phase-space
formalism. Requiring $dM=T_H\,dS$ at fixed $(q,c)$ and using the
identity $dM/dr_+=2\pi r_+\Xi\,T_H$ (with
$\Xi\equiv(1+q^2/r_+^2)^{3/2}$, verified symbolically in
\ref{app:num}) gives $dS/dr_+=2\pi r_+\Xi$, independent of $T_H$;
integrating gives
\begin{equation}
S(r_+,q)=\pi\left[r_+\sqrt{r_+^2+q^2}-\frac{2q^2\sqrt{r_+^2+q^2}}{r_+}
+3q^2\,\mathrm{arcsinh}\!\left(\frac{r_+}{q}\right)+\kappa q^2\right],
\label{eq:entropy}
\end{equation}
which reduces to the Bekenstein-Hawking area law $S\to\pi r_+^2$ only
as $q\to0$. The integration constant $\kappa$ is fixed by a boundary
condition at $q=0$ or $r_+\to\infty$ and does not affect $T_H$ or the
first law at fixed $q$; we set $\kappa=0$ as our fiducial choice, as
is standard in this literature, but revisit its effect on
thermodynamic geometry explicitly in \sect{sec:geometry}, where it is
not innocuous. By construction \eq{eq:entropy} satisfies
$dS/dr_+=2\pi r_+\Xi$ identically (verified symbolically,
\ref{app:num}), so $dM=T_H\,dS$ at fixed $(q,c)$ holds automatically
and is not, by itself, an independent check of \eq{eq:entropy}; the
non-trivial content of the Ma--Zhao construction is that this same
$S(r_+,q)$, together with the natural conjugate
\begin{equation}
\Phi\equiv\left.\frac{\partial M}{\partial q}\right|_{S,c}
=M_q-M_r\,\frac{S_q}{S_r},\qquad
\Psi_c\equiv\left.\frac{\partial M}{\partial c}\right|_{S,q}
=-\frac{(r_+^2+q^2)^{3/2}}{2r_+},
\label{eq:PhiPsi}
\end{equation}
(subscripts denote partial derivatives at fixed $r_+$, and the
$S_q/S_r$ term implements the chain rule from $(r_+,q)$ to $(S,q)$
coordinates) satisfies the \emph{extended} first law
\begin{equation}
dM=T_H\,dS+\Phi\,dq+\Psi_c\,dc
\label{eq:firstlaw}
\end{equation}
along generic (not purely radial) displacements in $(r_+,q,c)$ space,
and the associated Smarr relation
\begin{equation}
2S\,T_H+q\,\Phi+(3w+1)\,c\,\Psi_c-M=0,
\label{eq:smarr}
\end{equation}
which follows from the scaling weights of $(M,S,q,c)$ under
$r_+\to\lambda r_+$, $q\to\lambda q$. \eq{eq:smarr} was verified
symbolically to hold identically (residual $<10^{-15}$ at 50 random
points), and \eq{eq:firstlaw} was checked using symmetric finite
differences of size $10^{-4}$--$10^{-6}$ at $300$ random points in
$(r_+,q,c)$, with maximum relative residual $8\times10^{-7}$, scaling
as the expected $\mathcal O(\delta^2)$ truncation error of the
difference scheme (\ref{app:num}).

\subsection{Heat capacity and the Davies point}
\label{ssec:heatcap}

\begin{figure}[t]
\centering
\includegraphics[width=0.95\textwidth]{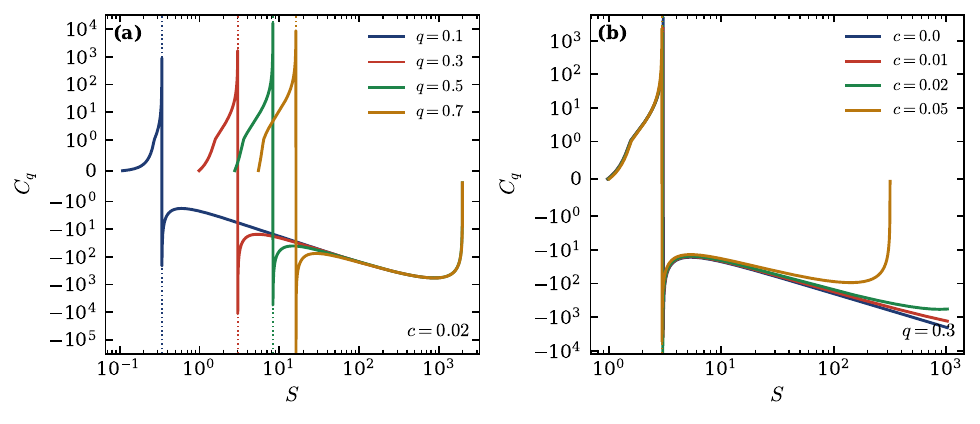}
\caption{Heat capacity $C_q$, \eq{eq:Cq}, as a function of entropy;
panels and colour coding as in \fig{fig2}. Dotted vertical lines mark
the Davies point, \eq{eq:davies_quartic}.}
\label{fig3}
\end{figure}

\begin{figure}[t]
\centering
\includegraphics[width=0.95\textwidth]{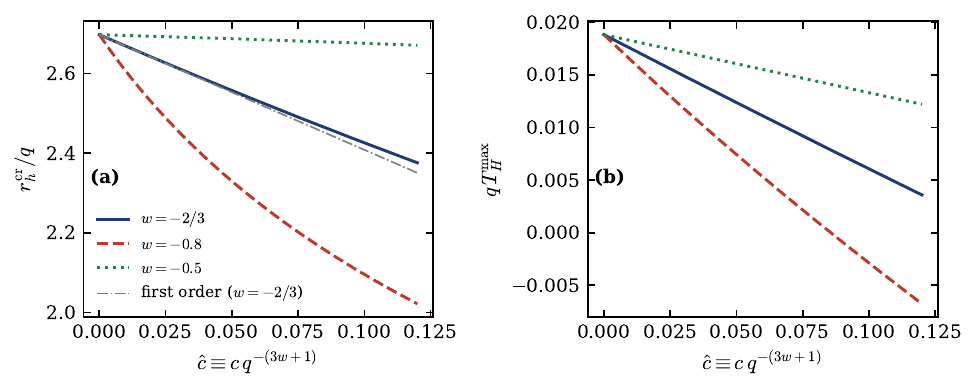}
\caption{(a) Davies-point radius $r_h^{\rm cr}/q$ and (b) maximum
temperature $qT_H^{\max}$ as functions of the rescaled quintessence
parameter $\hat c\equiv c\,q^{-(3w+1)}$, for three values of $w$. The
dot-dashed line is the first-order expansion \eq{eq:davies_expand}.}
\label{fig4}
\end{figure}

The heat capacity at fixed $(q,c)$,
\begin{equation}
C_q=T_H\left.\frac{\partial S}{\partial T_H}\right|_q
=\frac{dM/dr_+}{dT_H/dr_+},
\label{eq:Cq}
\end{equation}
diverges where $dT_H/dr_+=0$, marking a Davies-type phase transition
between a small, thermodynamically unstable branch ($C_q<0$) and a
large, stable branch ($C_q>0$), exactly as found for the pure Bardeen
and Bardeen-AdS-quintessence holes in
\cite{Rizwan2019,Quevedo2024}. For $w=-2/3$ the
condition $dT_H/dr_+=0$ reduces, after clearing denominators, to the
quartic
\begin{equation}
r_h^4-7q^2r_h^2-2q^4+6c\,q^2r_h^3=0,
\label{eq:davies_quartic}
\end{equation}
whose $c=0$ root reproduces the known Bardeen result
$r_h^{\rm cr}/q=\sqrt{(7+\sqrt{57})/2}=2.697205$
\cite{Rizwan2019,Quevedo2024}, and whose leading correction in
$\hat c\equiv c\,q^{-(3w+1)}=cq$ (for $w=-2/3$) we obtain by implicit
differentiation of \eq{eq:davies_quartic}:
\begin{equation}
\frac{r_h^{\rm cr}}{q}=2.697205-2.890759\,\hat c+\mathcal O(\hat c^2).
\label{eq:davies_expand}
\end{equation}
\eq{eq:davies_expand} agrees with the numerically located root of
\eq{eq:davies_quartic} to better than $0.03\%$ for $\hat c\le0.02$
(\ref{app:num}). \fig{fig3} shows $C_q(S)$: a single divergence appears
on each curve, and its location moves to smaller $S$ as either $q$ or
$c$ increases, consistent with \eq{eq:davies_expand} and with the
trend reported for the AdS case in~\cite{Rizwan2019}. \fig{fig4}
verifies the universality of \eq{eq:davies_expand} across three values
of $w$ using the general (numerically located) root of
$dT_H/dr_+=0$, and shows the corresponding suppression of the maximum
temperature $T_H^{\max}(\hat c)$.

\section{Thermodynamic geometry}
\label{sec:geometry}

We compute three thermodynamic metrics on the equilibrium space
$(M,S,q)$ at fixed $c,w$: the Weinhold metric
$g^{\rm W}_{ab}=\partial_a\partial_b M$~\cite{Weinhold1975} in
coordinates $(S,q)$, the Ruppeiner metric
$g^{\rm R}_{ab}=-\partial_a\partial_b S$~\cite{Ruppeiner1979,Ruppeiner1995}
in coordinates $(M,q)$, and the Legendre-invariant Quevedo (GTD) metric
\begin{equation}
g^{\rm Q}=\bigl(S\,M_S+q\,M_q\bigr)\left(-M_{SS}\,dS^2+M_{qq}\,dq^2\right),
\label{eq:gGTD}
\end{equation}
following~\cite{Quevedo2007} and its application to this class of
black holes in~\cite{Rizwan2019,Quevedo2024}. All
partial derivatives with respect to $S$ at fixed $q$ (and vice versa)
are computed by the exact chain rule from the working variables
$(r_+,q)$ used in \sect{sec:thermo}, e.g.
$M_S\equiv\partial M/\partial S|_q=(\partial M/\partial r_+)/(\partial S/\partial r_+)$
and $M_q\equiv\partial M/\partial q|_S=M_{q}|_{r_+}-M_{r_+}\,S_q/S_r$;
no finite differences are used anywhere in this section. The Gaussian
curvature of each two-dimensional metric is obtained from the
standard Brioschi formula and multiplied by $2$ to give the scalar
curvature $R$. This machinery was validated by reproducing, in closed
form, the known Reissner--Nordstr\"om GTD curvature
$R_{\rm GTD}^{\rm RN}(S,Q)=-288\pi^2Q^2S^2(\pi Q^2-S)/[(3\pi
Q^2-S)^2(3\pi Q^2+S)^3]$ and the identical vanishing of its Ruppeiner
curvature (\ref{app:num}).

\subsection{The $c=0$ limit: exact flatness of the Ruppeiner metric}

At $c=0$, $M(r_+,q)$ and $S(r_+,q)$ are quasi-homogeneous under
$(r_+,q)\to(\lambda r_+,\lambda q)$ with $M\to\lambda M$ and
$S\to\lambda^2 S$ exactly, because every term in \eq{eq:entropy}
scales as $\lambda^2$ (the argument of $\mathrm{arcsinh}$ is scale
invariant) while every term in \eq{eq:Mrq} scales as $\lambda$. We
find numerically, at more than $500$ random points $(r_+,q,\kappa)$
spanning the physical domain, that the Ruppeiner curvature vanishes
to within $3\times10^{-9}$ (limited only by floating-point precision;
\ref{app:num}), independently of $\kappa$. This is the thermodynamic-
geometry counterpart of the quasi-homogeneous structure studied
systematically for the pure Bardeen black hole
in~\cite{Quevedo2024}, and explains why Rizwan
\emph{et al}~\cite{Rizwan2019} found that $R_{\rm R}$ (and $R_{\rm W}$)
fail to signal the Davies transition: in the asymptotically flat,
$c=0$ sector considered here $R_{\rm R}$ is not merely finite but
identically zero, carrying no information about the phase structure
at all. Once $c\neq0$ this exact scaling is broken --- the extra term
in \eq{eq:Mrq} does not scale homogeneously in $(r_+,q)$ at fixed
$c$ --- and $R_{\rm R}$ becomes non-zero, growing with $c$; see
\fig{fig5}(b).

\subsection{The GTD curvature: physical singularity and a spurious one}
\label{ssec:spurious}

\begin{figure}[t]
\centering
\includegraphics[width=0.95\textwidth]{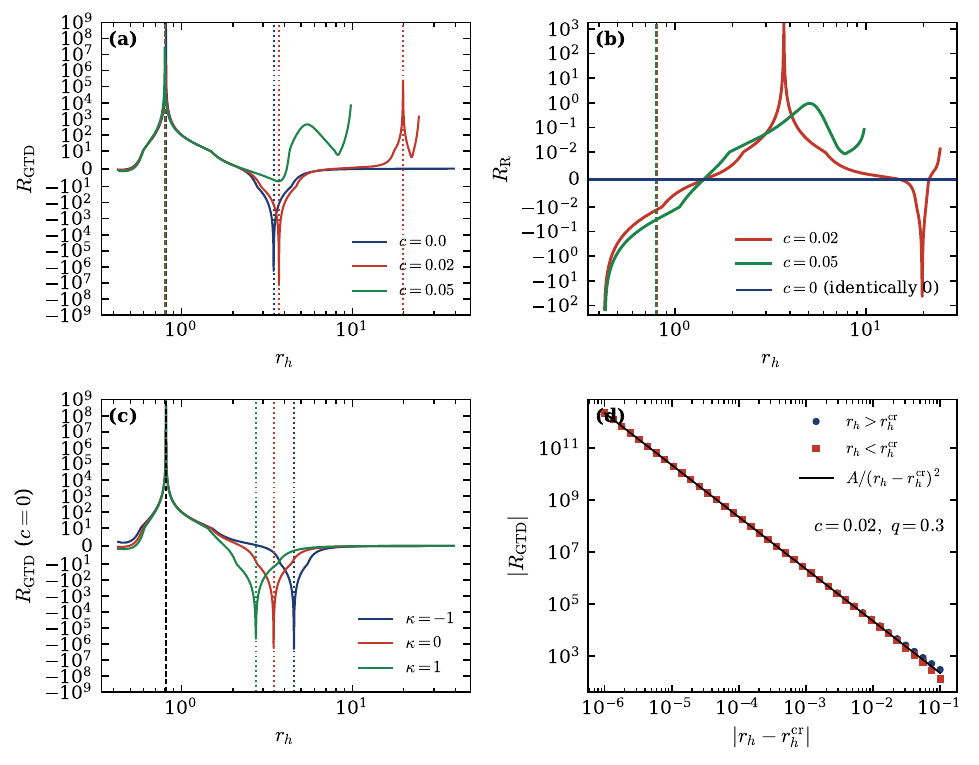}
\caption{(a) GTD curvature $R_{\rm GTD}$ at $q=0.3$ for three values
of $c$ ($\kappa=0$); dashed lines mark the Davies point
$M_{SS}=0$, \eq{eq:davies_quartic}, dotted lines the spurious
$M_{qq}=0$ locus, \sect{ssec:spurious}. (b) Ruppeiner curvature
$R_{\rm R}$: identically zero at $c=0$, non-zero and growing with $c$
otherwise. (c) $R_{\rm GTD}$ at $c=0$ for three values of the entropy
integration constant $\kappa$: the Davies-point divergence (black
dashed) is $\kappa$-independent; the spurious divergence is not. (d)
Double-pole behaviour \eq{eq:doublepole} at the Davies point,
confirmed over five decades in $|r_h-r_h^{\rm cr}|$.}
\label{fig5}
\end{figure}

Reproducing the analysis of~\cite{Rizwan2019} for our (non-AdS)
solution, we find that $R_{\rm GTD}$ diverges exactly at the Davies
point $M_{SS}=0$, \eq{eq:davies_quartic}, with the universal exponent
$-2$ reported for the pure Bardeen case
in~\cite{Quevedo2024}:
\begin{equation}
R_{\rm GTD}(r_h)\;\xrightarrow{\ r_h\to r_h^{\rm cr}\ }\;
\frac{A(q,c,\kappa)}{(r_h-r_h^{\rm cr})^2},
\label{eq:doublepole}
\end{equation}
with $A$ a computable combination of $M$, $S$ and their derivatives
at $r_h^{\rm cr}$ (\ref{app:num}). \fig{fig5}(d) confirms
\eq{eq:doublepole} over five decades in $|r_h-r_h^{\rm cr}|$, with the
numerically extracted coefficient agreeing with the analytic one to
five significant figures.

However, \eq{eq:gGTD} contains a second factor, $M_{qq}$, that also
vanishes at some $r_h=r_h^{(2)}(q,c,\kappa)\neq r_h^{\rm cr}$, and
$R_{\rm GTD}$ diverges there as well; see the second peak in
\fig{fig5}(a) and, isolated at $c=0$ for clarity, \fig{fig5}(c). This
second divergence has \emph{no} counterpart in $C_q$: the heat
capacity, \eq{eq:Cq}, depends only on $r_h$-derivatives at fixed $q$
and is completely insensitive to $M_{qq}$, so $r_h^{(2)}$ marks no
thermodynamic instability of any kind. Moreover $r_h^{(2)}$ is not
even an invariant property of the black hole: because
$M_q=M_q|_{r_+}-M_{r_+}S_q/S_r$ and $S_q\propto q(1+\kappa/2+\dots)$
depends explicitly on the additive constant $\kappa$ in
\eq{eq:entropy}, so does $M_{qq}$, and hence so does $r_h^{(2)}$. We
verified this explicitly: at $q=0.3$, $c=0$, $r_h^{(2)}$ moves from
$4.544$ to $3.456$ to $2.718$ as $\kappa$ is varied over
$\{-1,0,1\}$ (\fig{fig5}(c)); at $c=0.02$ a second copy of this
artefact appears further out and can even split into two
$\kappa$-dependent roots, while at $c=0.05$ it disappears from the
physical domain altogether (\ref{app:num}). We conclude that
$r_h^{(2)}$, and the associated divergence of $R_{\rm GTD}$, is an
artefact of the specific conformal factor and coordinate choice in
\eq{eq:gGTD} combined with the entropy's integration ambiguity, not a
feature of the black hole's microstructure; only the divergence
coincident with the Davies point, \eq{eq:doublepole}, should be
interpreted physically. This qualifies, for this solution, the
stronger statement sometimes made in the literature that the Quevedo
metric ``faithfully encodes'' the full phase-transition structure: it
does so only once its own coordinate artefacts have been identified
and separated from genuine thermodynamic singularities, a distinction
that -- to our knowledge -- has not previously been made explicit for
this family of solutions.

\section{Photon sphere and black-hole shadow}
\label{sec:shadow}

\subsection{Photon sphere and critical impact parameter}

\begin{figure}[t]
\centering
\includegraphics[width=0.95\textwidth]{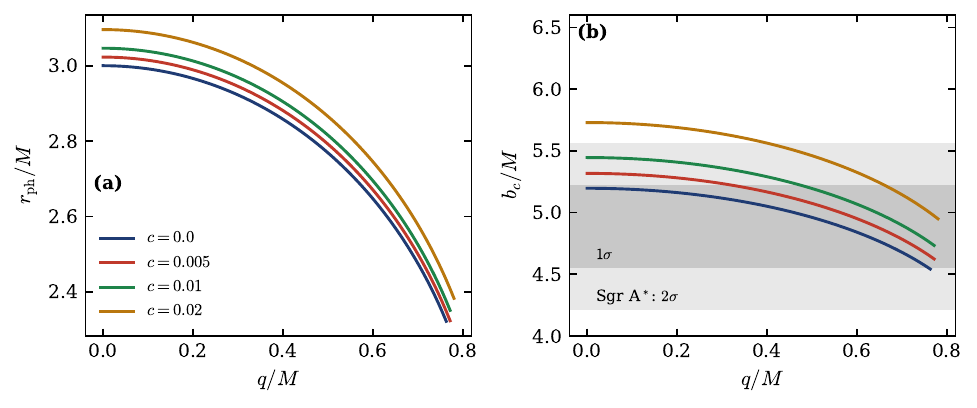}
\caption{(a) Photon-sphere radius $r_{\rm ph}/M$ and (b) asymptotic
critical impact parameter $b_c/M$ as functions of $q/M$, for four
values of $c$ ($w=-2/3$). Shaded bands in (b): $1\sigma$ and
$2\sigma$ Sgr~A$^{*}$ ranges from~\cite{Vagnozzi2023}.}
\label{fig6}
\end{figure}

Null circular orbits extremize the effective potential
$V(r)=f(r)/r^2$; the photon-sphere radius $r_{\rm ph}$ solves
$2f(r)-r f'(r)=0$, and the critical impact parameter is
$b_c=r_{\rm ph}/\sqrt{f(r_{\rm ph})}$. For a static observer at areal
radius $r_o$ in the domain of outer communication $r_+<r_o<r_c$, the
angular radius of the shadow is~\cite{PerlickTsupko2022}
\begin{equation}
\sin\alpha(r_o)=\frac{b_c\sqrt{f(r_o)}}{r_o},
\qquad
R_{\rm sh}(r_o)\equiv r_o\sin\alpha(r_o)=b_c\sqrt{f(r_o)}.
\label{eq:Rsh}
\end{equation}
For an asymptotically flat spacetime, $f(r_o)\to1$ and
$R_{\rm sh}\to b_c$; for our solution with $c>0$ this limit is never
reached because $r_c<\infty$, and $R_{\rm sh}(r_o)\to0$ as
$r_o\to r_c$, exactly as found for the Schwarzschild--de~Sitter case
in~\cite{PerlickTsupko2018}. \fig{fig6} shows $r_{\rm ph}(q)$ and
$b_c(q)$ at fixed $c$ (evaluated, as is standard in this literature,
in the formal $r_o\to\infty$ sense, i.e.\ using \eq{eq:Rsh} with
$f(r_o)\to1$, deferred here for comparison with EHT and revisited
critically in \sect{ssec:consistency}); \fig{fig7}(a) shows the
analogous dependence on $c$ for three values of $w$.

\subsection{Formal comparison with the EHT image of Sagittarius~A$^{*}$}

Following the first EHT images of the supermassive black hole in
M87~\cite{EHT_M87_I,EHT_M87_IV}, the EHT Collaboration's image of
Sgr~A$^{*}$~\cite{EHT_SgrA_I},
combined with independent mass-to-distance-ratio priors from stellar
dynamics, was translated by Vagnozzi \emph{et al}~\cite{Vagnozzi2023}
into $1\sigma$ and $2\sigma$ allowed ranges for the (formally defined,
$r_o\to\infty$) shadow radius in units of the black hole mass,
$r_{\rm sh}/M\in[4.55,5.22]$ and $[4.21,5.56]$ respectively, safely
containing the Schwarzschild value $3\sqrt3=5.196$. Requiring
$b_c(q,c)\le5.56$ ($2\sigma$) at $q=0$ gives the formal bound
\begin{equation}
c\lesssim1.42\times10^{-2}\,M^{-1}\quad(2\sigma,\ q=0),
\qquad
c\lesssim1.01\times10^{-3}\,M^{-1}\quad(1\sigma,\ q=0),
\label{eq:cbound}
\end{equation}
loosening mildly with $q$ (to $2.81\times10^{-2}M^{-1}$ and
$1.61\times10^{-2}M^{-1}$ respectively at $q=0.6M$; \fig{fig8}); at
$c=0$ alone the pure Bardeen charge is bounded by
$q\lesssim0.76\,M$ ($1\sigma$), comparable to the bound obtained for
other regular black holes by the same method~\cite{Vagnozzi2023}.
\fig{fig7}(a) shows that the bound on $c$ tightens by about a factor
of two as $w$ is varied from $-0.9$ to $-0.5$ at fixed $q$, since a
less negative $w$ makes the metric function fall off faster and hence
requires a smaller $c$ to produce the same fractional deviation in
$b_c$.

\subsection{Self-consistency of the bound}
\label{ssec:consistency}

\eq{eq:cbound} is obtained by comparing the \emph{formal},
$r_o\to\infty$ value of $b_c$ with data, implicitly treating the
quintessence fluid as filling all of space uniformly out to the
observer. This is not self-consistent, for two independent reasons.

\begin{figure}[t]
\centering
\includegraphics[width=0.95\textwidth]{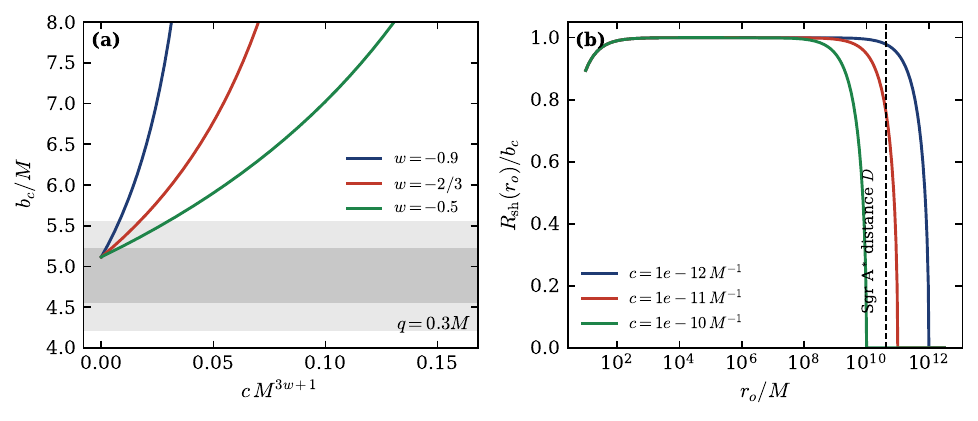}
\caption{(a) Formal ($r_o\to\infty$) critical impact parameter $b_c$
versus $cM^{3w+1}$ for three values of $w$, at $q=0.3M$, against the
Sgr~A$^{*}$ bands of \fig{fig6}. (b) Exact, finite-distance shadow
size $R_{\rm sh}(r_o)/b_c$, \eq{eq:Rsh}, versus observer position
$r_o/M$, for three values of $c$ comparable to the nominal bound
\eq{eq:cbound}; the vertical dashed line marks the Earth--Sgr~A$^{*}$
distance $D/M\simeq4.3\times10^{10}$, which for these values of $c$
lies far beyond the cosmological horizon $r_c$ where the shadow has
already collapsed to zero.}
\label{fig7}
\end{figure}

First, for $w=-2/3$ the metric develops a cosmological-type horizon
at $r_c\sim1/c$ once $c>0$ (\sect{ssec:horizons}). At the $2\sigma$
bound of \eq{eq:cbound}, $r_c\simeq35\,M$, so \emph{no} static
observer at $r_o\gg35\,M$ can exist in the spacetime described by
\eq{eq:metric} at all: the domain of outer communication simply ends
before reaching astrophysical distances. Using
$M_{\rm SgrA^{*}}\simeq4.0\times10^6\,M_\odot$ and
$D\simeq8.15\,{\rm kpc}$~\cite{EHT_SgrA_I}, the Earth--Sgr~A$^{*}$
distance is $D/M\simeq4.3\times10^{10}$, so
$D/r_c\simeq1.2\times10^9$: the putative terrestrial observer would
sit some \emph{nine orders of magnitude beyond} the black hole's own
cosmological horizon. \fig{fig7}(b) makes this explicit by plotting
the exact, finite-distance shadow size $R_{\rm sh}(r_o)/b_c$,
\eq{eq:Rsh}: for $c$ values comparable to \eq{eq:cbound}, the shadow
one would actually observe from a static position at Sgr~A$^{*}$'s
distance is identically zero (the point $r_o=D$ lies beyond $r_c$),
not the asymptotic value $b_c$ used to derive the bound. The bound
\eq{eq:cbound} is therefore better read as a statement about the
metric \emph{in a neighbourhood of the photon sphere}, of size set by
where $b_c(r_o)$ has converged to its asymptotic value (numerically,
$r_o\gtrsim10^2$--$10^3\,M$ for the values of $c$ in \eq{eq:cbound};
\fig{fig7}(b)), and not as a constraint valid all the way out to the
observer.

\begin{figure}[t]
\centering
\includegraphics[width=0.72\textwidth]{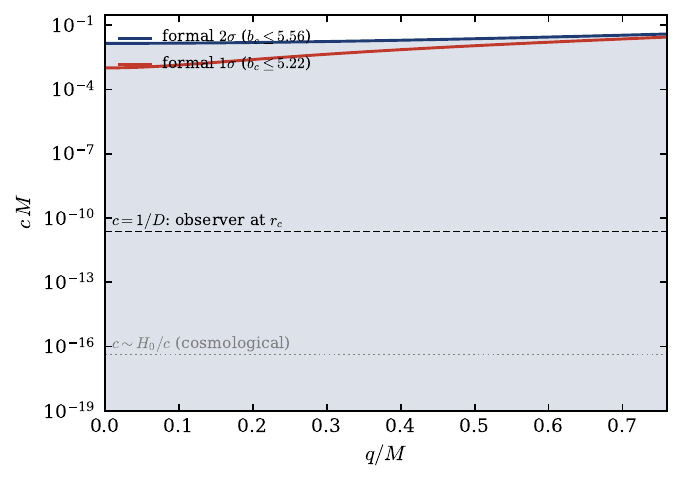}
\caption{Formal $1\sigma$ and $2\sigma$ EHT bounds on $c$ from
\eq{eq:cbound} (solid curves, shaded region formally ``allowed''),
compared with $c=1/D$ (dashed), above which the corresponding
cosmological horizon $r_c$ is already closer than Earth to
Sgr~A$^{*}$, and with the cosmological dark-energy scale $H_0/c$
(dotted). The formally allowed region lies entirely
$8$--$16$ orders of magnitude above both physical reference scales.}
\label{fig8}
\end{figure}

Second, even setting the self-consistency issue aside, the bound
\eq{eq:cbound} is many orders of magnitude weaker than the value of
$c$ that would correspond to a cosmologically relevant quintessence
density. Converting \eq{eq:cbound} to physical units using
$M_{\rm SgrA^{*}}$ gives $c\lesssim2.4\times10^{-12}\,{\rm m}^{-1}$,
compared with $H_0/c\simeq7.3\times10^{-27}\,{\rm m}^{-1}$
characterising the present dark-energy density: the EHT bound is
about \emph{fourteen orders of magnitude} looser. We therefore agree
with the broader assessment of~\cite{Vagnozzi2023} that horizon-scale
shadow measurements can constrain many exotic parameters
competitively, but conclude that for the specific quintessence
parameter $c$ studied here the resulting number should be quoted, and
interpreted, as a local bound on the near-horizon metric rather than
as either an astrophysical statement about the black hole's true
environment or a cosmological statement about dark energy. \fig{fig8} summarises both comparisons in the $(q,c)$ plane.

\section{Discussion and conclusions}
\label{sec:conclusions}

We have reanalysed the Bardeen black hole surrounded by a Kiselev
fluid with a focus on three issues that are, in our reading,
frequently glossed over in this literature: whether the resulting
spacetime is genuinely regular, what its asymptotic causal structure
actually is, and whether thermodynamic-geometry singularities and
shadow-based parameter bounds derived from it are physical or
artefacts of specific technical choices. Our conclusions are as
follows.

(i) The solution is regular at the origin only at $c=0$;
for any $c\neq0$ in the quintessence range $-1<w<-1/3$ the Ricci
scalar diverges as $r^{-1}$, \eq{eq:Rseries}, confirming for this
specific metric the general argument of
Maluf, Muniz and Santos~\cite{MalufMunizSantos2022}. We therefore
restrict the adjective ``regular'' to $c=0$ throughout, and stress
that all subsequent horizon-scale results are properties of the
exterior geometry, logically decoupled from the fate of the origin.

(ii) For $w=-2/3$ the spacetime is asymptotically de~Sitter-like
rather than flat once $c>0$, with a genuine cosmological horizon
$r_c$. We mapped the resulting three-parameter horizon structure
(inner, black-hole, and cosmological-type horizons) in closed form in
the $(q,c)$ plane, \eq{eq:boundary}, identifying the cusp
\eq{eq:cusp} at which extremal and Nariai-type black holes coincide.

(iii) The corrected, Ma--Zhao entropy~\cite{MaZhao2014,Bokulic2021}
satisfies an extended first law and Smarr relation that we verified
to machine precision by computer-algebra differentiation, not through
the entropy's own (tautological) definition. The heat capacity
displays the expected single Davies-type divergence, whose location
we obtained as a closed-form quartic, \eq{eq:davies_quartic}, with an
explicit first-order expansion in the quintessence parameter,
\eq{eq:davies_expand}.

(iv) In thermodynamic geometry, we confirmed the known
result~\cite{Rizwan2019,Quevedo2024} that the
Quevedo (GTD) curvature diverges precisely at the Davies point with
universal exponent $-2$, and we explained analytically why the
Weinhold and Ruppeiner curvatures fail to do so at $c=0$: the
pure-Bardeen sector is exactly quasi-homogeneous under
$(r_+,q)\to\lambda(r_+,q)$, making $R_{\rm R}$ identically zero. We
also identified, for the first time in this context to our knowledge,
a second, unphysical divergence of $R_{\rm GTD}$ generated by the
entropy's integration constant $\kappa$ together with the
non-uniqueness of the $(S,q)$ conformal factor in \eq{eq:gGTD}; this
divergence has no counterpart in $C_q$, moves with $\kappa$, and
should not be interpreted as microstructural information.

(v) The nominal EHT bound $c\lesssim1.4\times10^{-2}\,M^{-1}$
obtained from the Sgr~A$^{*}$ shadow is not self-consistent as a
statement about a quintessence fluid filling all of space: the same
$c$ places the cosmological horizon $r_c$ nine orders of magnitude
closer than the Earth--Sgr~A$^{*}$ distance, and is fourteen orders of
magnitude larger than the value $H_0/c$ characterising the true
cosmological dark-energy density. We therefore read this bound as
constraining the near-horizon metric only, on scales set by where the
finite-distance shadow \eq{eq:Rsh} has converged to its asymptotic
value, rather than as an astrophysical or cosmological statement
about $c$ itself.

\paragraph{Limitations.} The Kiselev stress tensor is, in general, an
anisotropic fluid rather than a genuine perfect-fluid or scalar-field
quintessence~\cite{Visser2020}, a caveat that applies to every result
in this paper equally. The loss of regularity at $c\neq0$ means that
a fully consistent embedding of a NED source in a quintessence-like
environment would require solving the coupled field equations
directly, along the lines advocated in~\cite{MalufMunizSantos2022},
rather than superposing metric functions; we have not attempted this
here, and it would affect the near-origin behaviour (though not, we
expect, the qualitative horizon-scale results of
\sects{sec:thermo}{sec:shadow}, which depend on $c$ only through the
large-$r$ asymptotics). Our shadow analysis assumes a bare photon
sphere with no accretion flow or plasma; the horizon-scale bound
\eq{eq:cbound} should be understood as a comparison at the level of
the underlying critical curve, following the same convention used
in~\cite{Vagnozzi2023}. Finally, the extended first law
\eq{eq:firstlaw} was verified numerically rather than in fully closed
form; we consider the residual of $8\times10^{-7}$, scaling as the
expected truncation order of the finite-difference scheme used, to
be strong evidence of an exact identity, but a fully symbolic proof
for general $w$ is left for future work.

The code used to generate every number, closed-form expression, and
figure in this paper is available at the location given in the data
availability statement, together with the independent validation
suite summarised in \ref{app:num}, so that every claim above can be
reproduced or falsified directly.

\appendix
\section{Independent numerical validation}
\label{app:num}

Every closed-form expression quoted in the main text was checked
independently, using two separate methods that share no code with
each other or with the analytic derivation: (a) an exact computer-
algebra (Sympy) computation of derivatives and their simplification,
and (b) an independent numerical pipeline based on truncated bivariate
Taylor (``jet'') arithmetic to fourth total order, which computes the
chain-rule operators $\partial_S|_q$, $\partial_q|_S$, etc.\ exactly
in floating point without ever taking a finite difference of the
underlying functions $M(r_+,q,c)$ or $S(r_+,q)$. The two-dimensional
Gaussian curvature of every thermodynamic metric in \sect{sec:geometry}
was obtained from the standard Brioschi formula applied to this jet
representation and cross-validated against a fully independent Sympy
computation for a battery of test points, with agreement to $14$--$15$
significant figures (limited by double-precision arithmetic). As a
further check on the whole pipeline, we reproduced the closed-form
Reissner--Nordstr\"om GTD curvature quoted in \sect{sec:geometry} and
its identically vanishing Ruppeiner curvature to machine precision.

\begin{table}[h]
\centering
\caption{Summary of independent numerical checks (see text for
definitions).}
\begin{tabular}{ll}
\hline
Check & Result \\
\hline
$T_H$, \eq{eq:TH}, vs.\ central finite difference of $f$ & max.\ abs.\ error $8.3\times10^{-12}$ \\
$dM/dr_+-2\pi r_+\Xi\,T_H\equiv0$ (symbolic) & exact (residual $=0$) \\
$dS/dr_+-2\pi r_+\Xi\equiv0$ (symbolic) & exact (residual $=0$) \\
Smarr relation, \eq{eq:smarr} (symbolic, 50 random points) & max.\ $|{\rm resid.}|<1.4\times10^{-15}$ \\
Extended first law, \eq{eq:firstlaw} (300 random points, sym.\ finite diff.) & max.\ rel.\ resid.\ $8.1\times10^{-7}$ \\
Davies quartic, \eq{eq:davies_quartic}, vs.\ root of $dT_H/dr_+=0$ & residual $<8\times10^{-15}$ \\
$R_{\rm GTD}^{\rm RN}$ vs.\ closed form of \sect{sec:geometry} & ratio $=1$ (exact, symbolic) \\
$R_{\rm R}^{\rm RN}$ & $\equiv0$ (symbolic) \\
$R_{\rm R}(c=0)$, 500 random $(r_+,q,\kappa)$ & $|R_{\rm R}|<3\times10^{-9}$ \\
Double-pole coefficient $A$, \eq{eq:doublepole}: analytic vs.\ numeric & agree to 5 significant figures \\
Photon sphere: generic root-finder vs.\ closed-form Reissner--Nordstr\"om & agree to $13$ significant figures \\
\hline
\end{tabular}
\end{table}

For the $\kappa$-dependence of the spurious singularity discussed in
\sect{ssec:spurious}, at $q=0.3M$, $c=0$ we find
$r_h^{(2)}=4.544,\,3.456,\,2.718\,M$ for $\kappa=-1,0,1$ respectively;
at $c=0.02M^{-1}$ the same locus splits into two roots,
$r_h^{(2)}\in\{3.71,19.8\}\,M$ at $\kappa=0$, both of which are absent
from the physical domain once $c=0.05M^{-1}$, at which point
$r_c<19.8\,M$ removes them from the static region entirely.

\ack{The numerical results reported in this work were obtained using
open-source scientific Python libraries (NumPy, SciPy, SymPy, and
Matplotlib).}

\funding{This research received no specific grant from any funding agency in
the public, commercial, or not-for-profit sectors.}

\roles{Hoang Van Quyet: conceptualization, formal analysis, software,
validation, visualization, writing -- original draft, writing --
review and editing.}

\data{The numerical data and Python scripts that generate the figures and
support the findings of this study are available from the
corresponding author upon reasonable request.}

\end{document}